\documentclass[manuscript]{rmaa}
\usepackage{paralist}
\usepackage{url}
\usepackage{multirow}
\usepackage{amsmath}

\usepackage{psfrag,color}
\usepackage[utf8]{inputenc}

\title{Instrumental Broadening of the SPOL spectropolarimeter at the University of Arizona} 

\author{
  Alfredo Amador-Portes\altaffilmark{1},
  Vahram Chavushyan\altaffilmark{1},
  Victor M. Patiño-Alvarez\altaffilmark{1,2}}

\altaffiltext{1}{Instituto Nacional de Astrofísica, Óptica y Electrónica, Luis Enrique Erro \#1, Tonantzintla, Puebla, México, C.P. 72840.}

\altaffiltext{2}{Max-Planck-Institut für Radioastronomie, Auf dem Hügel 69, D-53121 Bonn, Germany.}

\shortauthor{Amador-Portes, Chavushyan, \& Patiño-Alvarez}
\shorttitle{Instrumental Broadening at the SO}

\fulladdresses{
\item Alfredo Amador-Portes and Vahram Chavushyan: Instituto Nacional de Astrofísica, Óptica y Electrónica, Luis Enrique Erro \#1, Tonantzintla, Puebla, México, C.P. 72840 (aamador, vahram@inaoep.mx).
\item Victor M. Patiño-Alvarez: Instituto Nacional de Astrofísica, Óptica y Electrónica, Luis Enrique Erro \#1, Tonantzintla Puebla, México, C.P. 72840 (victorp@inaoep.mx), and Max-Planck-Institut für Radioastronomie, Auf dem Hügel 69, D-53121 Bonn, Germany.}

\listofauthors{Amador-Portes, Chavushyan, \& Patiño-Alvarez}
\indexauthor{Alfredo Amador-Portes.}
\indexauthor{Vahram Chavushyan.}
\indexauthor{Victor M. Patiño-Alvarez.}

\abstract{The Ground-based Observational Support of the Fermi Gamma-ray Space Telescope is conducted by the University of Arizona using the 2.3m Bok and 1.54m Kuiper telescopes operated by the Steward Observatory (SO). This program monitors blazar sources with spectroscopic (among others) observations. Yet, the instrumental broadenings for the different slit widths used in the spectra, are unavailable (the widths range from $2\farcs0$ to $12\farcs7$.). Using quasi-simultaneous spectroscopic observations of the blazar 3C 273 between the SO, Observatorio Astrofísico Guillermo Haro (OAGH), and Observatorio Astronómico Nacional San Pedro Mártir (OAN-SPM), we can provide an estimation of the instrumental profile for two of the slit widths. Since the instrumental broadening and the slit width are directly proportional, we were able to estimate the instrumental broadening for all six slit widths used at the SO. We found significant variations between the instrumental profiles, emphasizing the need to correct the instrumental broadening for each aperture.}

\resumen{El soporte de observaciones terrestres del telescopio espacial Fermi de rayos gamma, es realizado por la Universidad de Arizona usando los telescopios Bok de 2.3m y Kuiper de 1.54m, operado por el Observatorio Steward (SO). Este programa monitorea fuentes blazares con observaciones espectroscópicas (entre otras). Sin embargo, la información del perfil instrumental para diferentes anchos de rendija usados en los espectros ópticos no está disponible (los anchos van desde $2\farcs0$ a $12\farcs7$). Por lo que utilizando observaciones de espectros cuasi-simultáneos del blazar 3C 273 entre el SO, el Observatorio Astrofísico Guillermo Haro (OAGH) y el Observatorio Astronómico Nacional San Pedro Mártir (OAN-SPM), se pudo determinar el ensanchamiento instrumental para 2 anchos de rendija. Ya que el ancho instrumental y el ancho de rendija son directamente proporcionales, fue posible estimar el ancho instrumental para los seis anchos de rendija usados en el SO. Encontramos variaciones significativas entre los perfiles instrumentales, lo que enfatiza la necesidad de corregir el ensanchamiento instrumental para cada apertura.}

\addkeyword{Galaxies: Active}
\addkeyword{Instrumental Broadening}
\addkeyword{Quasars: Individual: 3C 273}
\addkeyword{Steward Observatory: SPOL}

\begin{document}
\maketitle

\section{Introduction}
\label{sec:introduction}

Since the launch of the Fermi Gamma-Ray Space Telescope, there have been counterpart supporting monitoring programs in the entire electromagnetic spectrum, from radio waves to X-rays (e.g. SMA, \citealp{SMAmonitoring}; OVRO, \citealp{OVROmonitoring}, SMARTS; \citealp{SMARTSmonitoring}, Swift; \citealp{SWIFTmonitoring}). In particular, we focus on the Ground-based Observational Support of the Fermi Gamma-ray Space Telescope carried out at the University of Arizona using the 2.3m Bok (Kitt Peak) and 1.54m Kuiper (Mt. Bigelow) telescopes operated by the Steward Observatory (SO) \citep{SmithSO2009}\footnote{2009 Fermi Symposium, eConf Proceedings C091122}. This is one of the most important monitoring programs because it is the most complete public database of optical spectroscopy, photometry, and polarimetry, for blazars, with the most number of sources and best cadence of observation\footnote{\url{http://james.as.arizona.edu/~psmith/Fermi/}}. In this program, the spectroscopic observations are conducted with the SPOL spectropolarimeter which uses up to 6 configurations (apertures) corresponding to six different slit widths (1=2\farcs0, 2=3\farcs0, 3=4\farcs1, 4=5\farcs1, 5=7\farcs6, and 6=12\farcs7). \citet{SmithSO2009} also provides a spectral resolution range of $15-25\text{ \AA}$, depending on the slit width chosen for the observation.

Unfortunately, the instrumental broadening for each slit width is not publicly available, this has prevented us from studying the variability of the FWHM of the broad emission lines present in Flat-Spectrum Radio Quasars (FSRQ), unlike the study of flux variability previously done in various works for the Mg II $\lambda2798$ \citep{Leon2013, Chavushyan2020, Antonio2021}, H$\beta$ \citep{Fernandes2020}, and C IV $\lambda1549$ \citep{Antonio2022} emission lines. 

FSRQ are a sub-type of blazars \citep{Urry1995} characterized by their high variability across the entire electromagnetic spectrum (e.g. \citealp{Aharonian2007, Antonio2021}), with variations on very different time scales, even in the same wavelength range (e.g. \citealp{Fan2018, Gupta2018}). When there is a flux increase well above the average emission, in a very short time, we call these phenomena a flare; among AGNs, these are mostly found in blazars. There is observational evidence that the emission line flares in some blazars can be driven by non-thermal emission from the jet (e.g. \citealp{Leon2013, Chavushyan2020, Antonio2021, Antonio2022}), so it is interesting to study if the emission line profile also changes during these flares. Such a study could involve a detailed analysis of emission lines during both quiescent and flaring states, aiming to characterize any changes in emission line profiles, including shifts in peak wavelength, changes in line width (FWHM), and alterations in profile asymmetry.

The importance of monitoring the FWHM is that the emission lines provide information about the kinematics and dynamics of the gas in the Broad-Line Region (BLR) surrounding the central supermassive black hole (SMBH). Therefore, understanding the characteristics of the BLR contributes to the knowledge of accretion processes, and gravitational potential. For example, SMBH mass estimation techniques like reverberation mapping \citep{ReverberationBlandford, ReverberationPeterson}, and single-epoch spectra \citep{GreeneMass, KongMass, VestergaardMass, ShawMass}, directly use this parameter.

It is clear that obtaining and understanding the instrumental line profile is a fundamental step in studying emission line variability in AGNs and other astronomical sources, as it enables the separation of the intrinsic characteristics of the source from instrumental effects, leading to more accurate scientific conclusions. The observed profile ($\text{FWHM}_{obs}$) is a convolution of both, the intrinsic profile of the source ($\text{FWHM}_{corr}$) and the instrumental profile ($\text{FWHM}_{inst}$, also known as the spectral resolution). This convolution is typically represented mathematically as a quadratic sum, shown in Equation \ref{eq:sum}.

\begin{equation}
\text{FWHM}_{obs}^{2}=\text{FWHM}_{corr}^{2}+\text{FWHM}_{inst}^{2}
\label{eq:sum}
\end{equation}

\citet{Nalewajko3C454.3} studied line width behavior without performing correction for instrumental broadening, which yielded an inconclusive result. This fact led to the establishment of techniques to counteract the lack of knowledge of instrumental broadening for SO spectroscopic data. \citet{Zhang20193C273} used a single value for $\text{FWHM}_{inst}$ estimated from other spectroscopic observations (not quasi-simultaneous) to apply it to observations with different slit widths. \citet{Rakshit2020PKS} used the same value and applied it to the SO spectra with different slit widths. The latter can lead to underestimation or overestimation of intrinsic emission line width in a set of spectra with different slit widths. Even using only spectra taken with a single slit width as in \citet{PandeyPKS0736}, FWHM measurements of emission lines will be overestimated without instrumental broadening correction. However, studies of the relative line width behavior can still be done in this case.

A proper approach to this problem would be to use calibration lamps or night sky lines to measure the instrumental profile. Unfortunately, as these original data were not available, we had to use instead a different method. Use spectra from the same source from an observatory with known instrumental widths, selecting only observations quasi-simultaneous to the SO data  (within 24 hours of each other), and taking into account the possible intra-day variability (IDV) that Blazars can show \citep{Gupta2018}. Assuming that the corresponding intrinsic line profile is the same in the spectra from both observatories, we can use the quasi-simultaneous data to estimate the SO instrumental profile. This procedure must be performed for each slit width because they have different instrumental broadening. 

The spectrograph resolution depends on a series of factors that we assume remain unchanged in the observations made in the SO, with the exception of the slit width. Hence, the spectral resolution (instrumental broadening) should be proportional to this slit width\footnote{\url{http://james.as.arizona.edu/~psmith/SPOL/gratings.html}} ($\text{FWHM}_{inst}\propto\text{SW}$, e.g. \citealp{1974apoi.book..463S}). Also, the ratio between the widths of the slits will reflect the expected difference in instrumental broadening.

\section{Observational Data}

We search for quasi-simultaneous observations between the SO, the OAGH operated by the Instituto Nacional de Astrofísica, Óptica y Electrónica\footnote{\url{https://astro.inaoep.mx/observatorios/oagh/}} (INAOE) and the OAN-SPM operated by the Universidad Nacional Autónoma de México\footnote{\url{https://www.astrossp.unam.mx/es/}} (UNAM), in which for the last two we have knowledge of the instrumental profile used in every observation. The instrument used in the OAGH and OAN-SPM observations is the Boller \& Chivens spectrograph. The best candidate for this analysis was the FSRQ blazar 3C 273, due to the high number of observations, high S/N ratio in the spectral range between 4000-7000 Å, it does not present telluric absorption lines around the emission lines of interest, and finally, its redshift (0.158) allows for the presence of a narrow emission line ([O III] $\lambda5007\:\text{\AA}$), which in AGNs, is not expected to vary significantly over a long period of time (e.g. \citealp{Shapovalova3C390.3}).

We made a further filter in the spectra set for all observatories by removing spectra with a low S/N ratio. We end up with 5 spectra taken with a slit width of 5.1 arcsec (SO), in which we have quasi-simultaneous data in the OAGH or OAN-SPM, while there are 19 spectra taken with a slit width of 7.6 arcsec (SO), with quasi-simultaneous data.

\section{Data Analysis}
As mentioned before, for the OAGH and OAN-SPM spectra the instrumental broadening is known, because, for each object spectra, we take a comparison lamp of He-Ar (OAGH) and Cu-Ne-He-Ar (OAN-SPM) that is used for wavelength calibration. Therefore the spectrum of the comparison lamp uses the same setup and light path as the spectrum of the blazar, this justifies using the width of the lines in the comparison lamp as the instrumental broadening for the object spectra correction. Hence, for each lamp spectrum, the FWHM of the non-blended lines was measured. We estimated a mean value, with its error as the quadratic sum of the standard deviation of the measurements and the dispersion of the spectrum (\AA/pixel).

\begin{figure}[h!]
\centering
\includegraphics[width=\columnwidth]{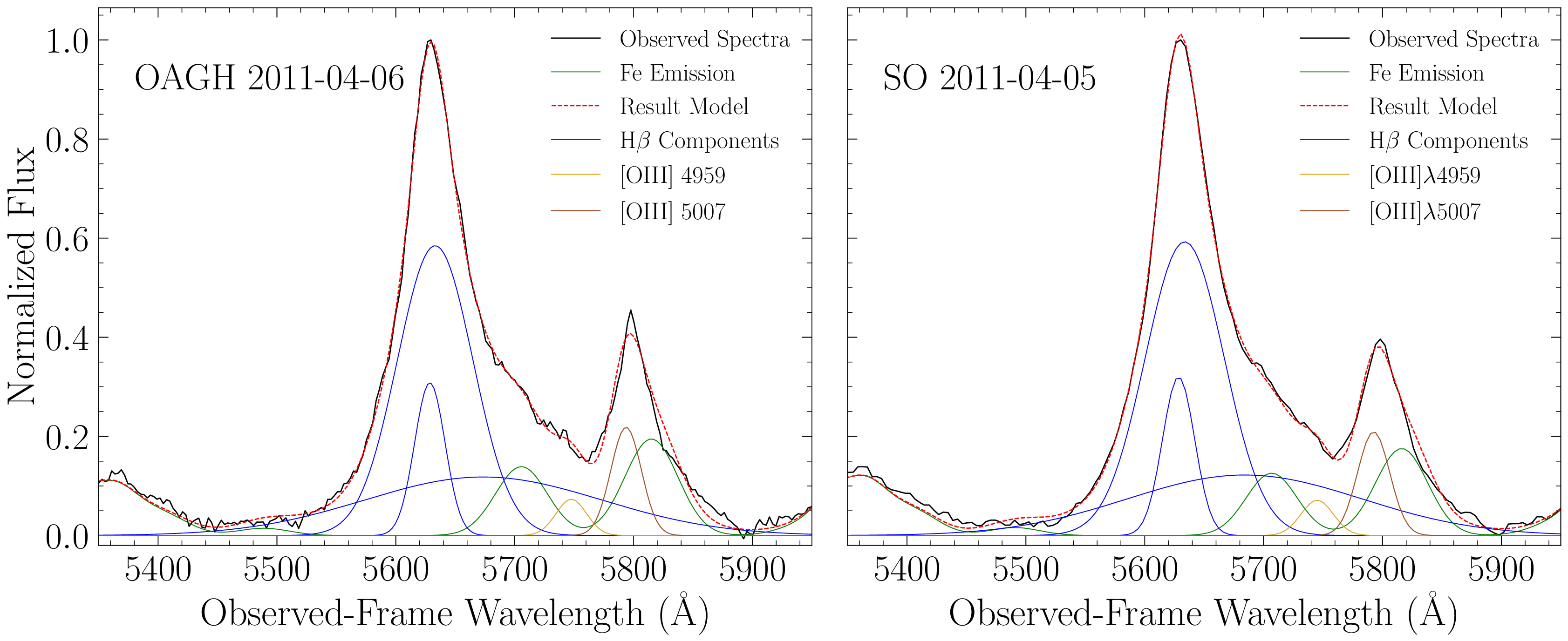}
\caption{Example of spectral decomposition for the H$\beta$ region in the 3C 273 data (observed frame) Left Panel: Spectra from the SO database. Right Panel: Spectra from OAGH. The spectra were taken within 24 hours of each other.}
\label{fig:spectra}
\end{figure}

For all the spectra, we left them in the observed frame because the instrumental broadening is geometrical, i.e. a fixed value in pixels. We then fitted a local continuum and subtracted it from the spectra. The spectral features were fitted with Gaussian functions in the range from $5500-5900\text{ \AA}$ which correspond to the H$\beta$ region in the observed frame, and consequently, the emission lines fitted were H$\beta$ (with narrow, broad, and very-broad components), [O III]$\lambda\lambda4959,5007\text{ \AA}$ (with one Gaussian each), and the Fe II emission. The same multi-Gaussian model was used in each pair of spectrum, allowing to vary only the FWHM and amplitude. An example can be seen in Figure \ref{fig:spectra} where the pair of spectra are normalized to the maximum flux of the H$\beta$ emission.

All measurements in this paper were made using the [O III] $\lambda5007$ \AA\ emission line, in the observed frame; which at the redshift of 3C 273, the line is at 5798 \AA. This line is produced in the Narrow-Line Region (NLR), and due to its large size, they are not expected to change over large periods of time because the ionizing radiation will respond more slowly (\citealp{Shapovalova3C390.3}, and references therein). Previous studies of IDV on 3C 273 mainly focus on high-energy and optical bands and have not been reported for narrow lines. However, as mentioned earlier we do not expect such fast variations in narrow emission lines. An example of the low expected variations in this line is presented in \citet{Yuan20223C273}, where they calculated [O III] $\lambda5007$ \AA\ emission line flux. The variation between each day was always less than the uncertainties that could be obtained from the spectra. Additionally for each pair of spectra, if the multi-Gaussian model was not fitted with similar amplitudes (taking into account that the width changes), then the spectra pair was discarded to avoid possible IDV.

From all the measurements obtained, we estimate the intrinsic profile and their uncertainty with error propagation for each quasi-simultaneous observation. Assuming that the intrinsic profiles for each date on the spectra of both observatories are the same (since they are quasi-simultaneous), we were able to estimate the instrumental broadening of the SO spectra, using the intrinsic profile widths obtained from OAGH and OAN-SPM, and the observed profile widths measured in the SO spectra. We estimated a weighted mean for the instrumental broadening measurements of the two slit widths we have available and calculated the standard error of the weighted mean.

\section{Results}

This analysis shows us that the instrumental profile for the SO spectra for the slit widths of 5.1, and 7.6 arcsec are $16.55\pm4.43$, and $23.23\pm1.79\text{ \AA}$, respectively. Given that the instrumental broadening should be proportional to the slit width, we can take any of the instrumental broadening measurements we have and obtain expected values for the instrumental profiles of the other slit widths. This means that for any two slit widths (e.g. $a$ and $b$) these values must satisfy $\text{FWHM}_{inst}(a)/\text{FWHM}_{inst}(b)=\text{SW}(a)/\text{SW}(b)$. For example, for the slit width of 4.1 (that we do not measure) and 7.6 (that we do measure), the instrumental broadening for the former is calculated as $\text{FWHM}_{4.1}=\text{FWHM}_{7.6}\times4.1/7.6$.

\begin{figure}[h!]
\centering
\includegraphics[width=\columnwidth]{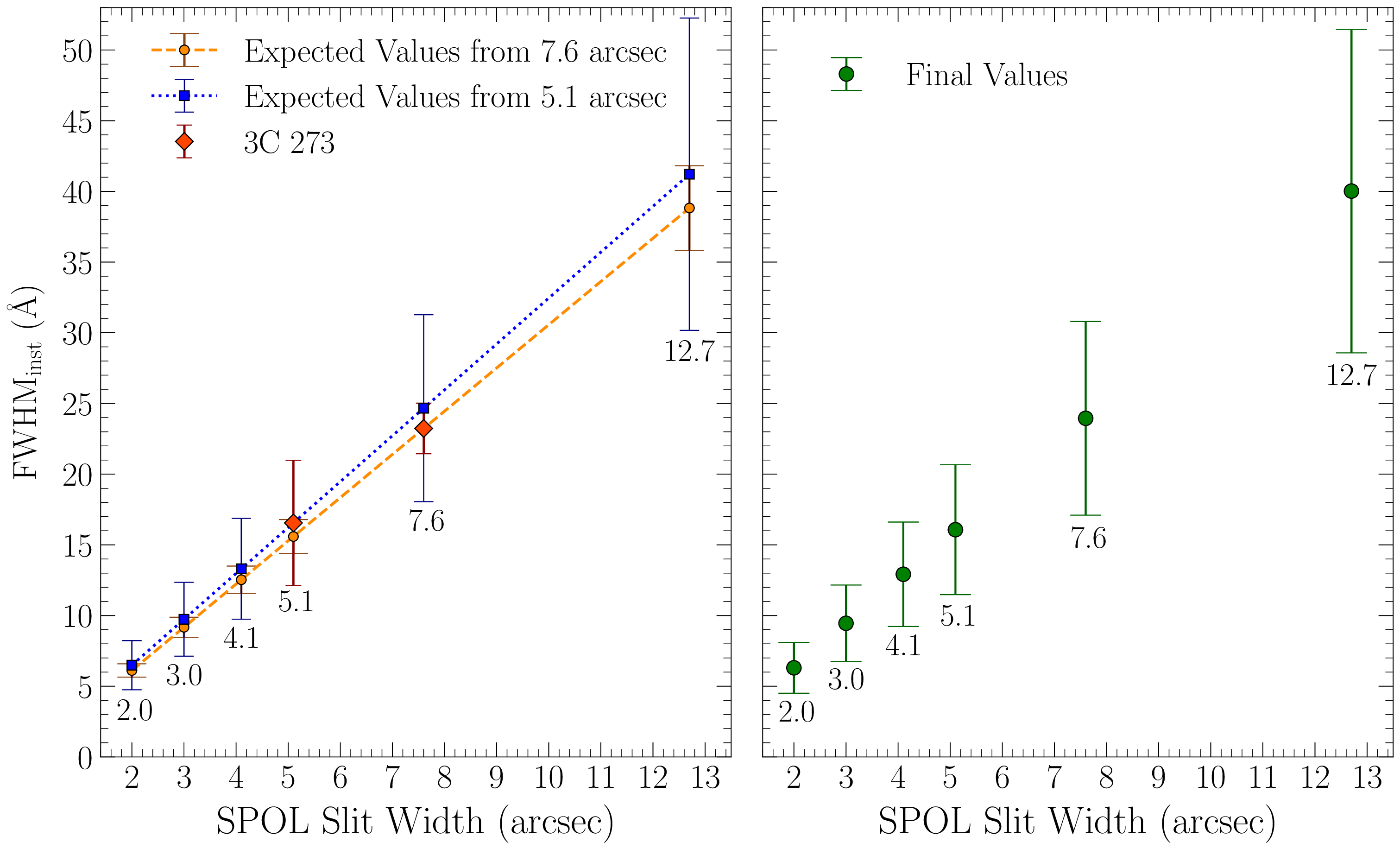}
\caption{Left Panel: Expected sets of instrumental profiles with their uncertainties, measured using a 5.1 arcsec (navy blue squares and dotted line), and a 7.6 arcsec (orange dots and dashed line) slit width as a reference point. Observational measurements of instrumental broadening are also shown (red diamonds). Right Panel: Instrumental broadening estimations for all 6 slit widths based upon the mean of the two expected values for each slit width.}
\label{fig:profiles}
\end{figure}

Since we have two measurements of the instrumental broadening, we obtained two sets of expected instrumental profiles, with their uncertainties, for all the slit widths. The difference between the values obtained for each slit width is most likely due to the uncertainties. We then calculated the mean of these two values for each slit width as well as its errors. These different values of instrumental broadening for each slit width are presented in Table \ref{tab:profiles}.

In Figure \ref{fig:profiles} left panel we present the instrumental broadening obtained through this analysis for the spectra taken with slit widths of 5.1 and 7.6 (red diamonds), and the expected values estimated with both measurements (blue squares and orange circles respectively). The final instrumental broadening values for all the slit widths (green circles) are presented in Figure \ref{fig:profiles} right panel.

\begin{table*}[h!]
\centering
\newcommand{\DS}{\hspace{6\tabcolsep}} 
\caption{I\MakeLowercase{nstrumental broadening for each slit width of the} SPOL \MakeLowercase{at the} SO.}
\label{tab:profiles}
\setlength{\tabnotewidth}{\linewidth}
\setlength{\tabcolsep}{1.2\tabcolsep} \tablecols{5}
\begin{tabular}{cccc}
\toprule
 & \multicolumn{3}{c}{ Instrumental Broadening (\AA)}\\
\cmidrule{2-4}
\multirow{2}{5em}{\centering Slit Width (arcsec)} & \multicolumn{2}{c}{Expected Values} & \multirow{2}{7em}{\centering Final Values} \\
\cmidrule{2-3}
 & From 5.1 arcsec & From 7.6 arcsec &  \\
\midrule
2.0 & $6.49\pm1.74$ & $6.11\pm0.47$ & $6.30\pm1.80$ \\
3.0 & $9.74\pm2.61$ & $9.17\pm0.71$ & $9.45\pm2.70$ \\
4.1 & $13.31\pm3.57$ & $12.53\pm0.97$ & $12.92\pm3.69$ \\
5.1 & $16.55\pm4.44$ & $15.59\pm1.20$ & $16.07\pm4.59$ \\
7.6 & $24.67\pm6.61$ & $23.23\pm1.79$  & $23.95\pm6.85$ \\
12.7 & $41.22\pm11.04$ & $38.82\pm2.99$ & $40.02\pm11.44$ \\
\bottomrule
\end{tabular}
\end{table*}

\section{Conclusions} \label{sec:conclusions}

We have estimated the instrumental broadening for the six different slit widths used in the spectroscopic observations carried out by the Steward Observatory\footnote{\url{http://james.as.arizona.edu/~psmith/Fermi/}}, and the results are presented in Table \ref{tab:profiles}. Notably, there is a significant difference in instrumental broadening across the different slit widths (which is expected), with the ratio between the maximum and minimum broadening being approximately 6.35 times. This highlights the importance of considering distinct instrumental broadening values for each slit width. Using a fixed or mean instrumental broadening for all slit widths will lead to an overestimation for smaller slit widths and an underestimation for the larger ones. Even when using mean and root mean square (rms) spectra with multiple slit widths, it is not correct to use a single instrumental broadening value, because the resulting spectra would still retain broadening information from all the slit widths, and it is strongly dependent on the most used setup.

\section{Acknowledgments}

We thank the anonymous referee for the constructive comments that helped to improve the manuscript. A.A.-P. acknowledges support from the CONAHCYT (Consejo Nacional de Humanidades, Ciencia y Tecnología) program for Ph.D. studies. This work was supported by CONAHCYT research grants 280789 and 320987. In addition, this work was supported by the Max Plank Institute for Radioastronomy (MPIfR) - Mexico Max Planck Partner Group led by V.M.P.-A. Data from the Steward Observatory spectropolarimetric monitoring project were used. This program is supported by Fermi Guest Investigator grants NNX08AW56G, NNX09AU10G, NNX12AO93G, and NNX15AU81G. This publication is based on data collected at the Observatorio Astrofísico Guillermo Haro (OAGH), Cananea, Sonora, Mexico, operated by the Instituto Nacional de Astrofísica, Óptica y Electrónica (INAOE). Funding for the OAGH has been provided by CONAHCYT. This paper is based upon observations carried out at the Observatorio Astronómico Nacional on the Sierra San Pedro Mártir (OAN-SPM), Baja California, México.

\end{document}